\documentclass[10pt, conference]{IEEEtran}
\usepackage{amsmath,amssymb}
\usepackage[linesnumbered,ruled]{algorithm2e}
\usepackage{multirow}
\usepackage{graphicx,xcolor,array}
\usepackage{pgfplots}
\usepackage{tikz}
\usepackage{flushend}
 \flushend
\usetikzlibrary{fit,positioning,calc, backgrounds}
\usetikzlibrary{shadows,hobby}
\usetikzlibrary{fadings}
\usetikzlibrary{shapes.arrows,calc,quotes,babel}
\usetikzlibrary{graphs,graphs.standard,arrows.meta, shapes.misc, positioning,decorations.pathreplacing,calligraphy}
\pgfplotsset{compat=1.14}
\usepackage{tikz,cite}
\usetikzlibrary{arrows}
\usetikzlibrary{automata,positioning}
\usepackage{makecell}
\usepackage{upgreek}
\usepackage{pifont}
\usepackage{mathscinet}
\usepackage[letterpaper, left=0.625in,right=0.625in,top=0.75in,bottom=1in]{geometry}
\usepackage{geometry}
\usepackage[flushleft]{threeparttable}
\usepackage{enumitem}
\usepackage[colorlinks=true,linkcolor=black,anchorcolor=black,citecolor=black,filecolor=black,menucolor=black,runcolor=black,bookmarks=false,hidelinks,urlcolor=black]{hyperref}
\usepackage{enumitem}

\newif\ifcomment
\commenttrue

\makeatletter
\def\endthebibliography{%
  \def\@noitemerr{\@latex@warning{Empty `thebibliography' environment}}%
  \endlist
}
\makeatother

\begin{document}
\title{SoK: DAG-based Consensus Protocols}
 \author{
 \IEEEauthorblockN{
 Mayank Raikwar,\IEEEauthorrefmark{1} Nikita Polyanskii,\IEEEauthorrefmark{2} Sebastian M{\"u}ller\IEEEauthorrefmark{3}
 \IEEEauthorblockA{\IEEEauthorrefmark{1}University of Oslo, Norway\\
 Email: mayankr@ifi.uio.no}
 \IEEEauthorblockA{\IEEEauthorrefmark{2}IOTA Foundation, Berlin, Germany\\
 Email: nikita.polyansky@gmail.com}
 \IEEEauthorblockA{\IEEEauthorrefmark{3}Aix Marseille Universit{\'e}, CNRS, Centrale Marseille, France\\
 Email: sebastian.muller@univ-amu.fr}
 }}

\pagestyle{empty}
\maketitle

\begin{abstract}
This paper is a Systematization of Knowledge (SoK) on Directed Acyclic Graph (DAG)-based consensus protocols, analyzing their performance and trade-offs within the framework of consistency, availability, and partition tolerance inspired by the CAP theorem.

We classify DAG-based consensus protocols into availability-focused and consistency-focused categories, exploring their design principles, core functionalities, and associated trade-offs. Furthermore, we examine key properties, attack vectors, and recent developments, providing insights into security, scalability, and fairness challenges. Finally, we identify research gaps and outline directions for advancing DAG-based consensus mechanisms.
\end{abstract}

\section{Introduction}

Distributed Ledger Technology (DLT) has become fundamental in supporting secure and transparent transaction systems across decentralized networks. Maintaining an immutable and append-only ledger, DLT enables a trustless environment where participants can transact directly without intermediaries. This technology, particularly its application in cryptocurrencies like Bitcoin and Ethereum, has attracted wide interest due to its potential for enhanced transparency, operational efficiency, and decentralization. However, standard blockchain architectures face key performance limitations, including low throughput and high confirmation latency. Studies have further highlighted trade-offs in speed, security~\cite{kiayias2015speed}, and performance~\cite{wu2023security}, as well as the inherent challenge in achieving decentralization, consistency, and scalability simultaneously (the DCS-satisfiability theorem)~\cite{trilemma, kalajdjieski2022databases}. In response, DAG-based DLTs have emerged, using alternative consensus structures to improve scalability and support more efficient consensus mechanisms.

\subsection{What is a DAG-based Consensus Protocol}

Traditional blockchain protocols, such as Bitcoin’s Nakamoto consensus~\cite{nakamoto2008bitcoin}, arrange blocks of transactions sequentially in a chain, where each block references its predecessor, extending back to the genesis block. Consensus in these systems often proceeds in rounds, with new blocks added to the longest chain, which participants recognize as the valid ledger. This linear structure, however, imposes limitations on scalability and throughput.

In contrast, DAG-based consensus protocols allow blocks to reference multiple predecessors, creating a Directed Acyclic Graph (DAG) structure. This referencing mechanism enables a framework where multiple blocks can be added concurrently, supporting parallel processing of transactions. By diverging from the sequential constraints of traditional models, DAG-based protocols offer improved scalability and flexibility in transaction processing.

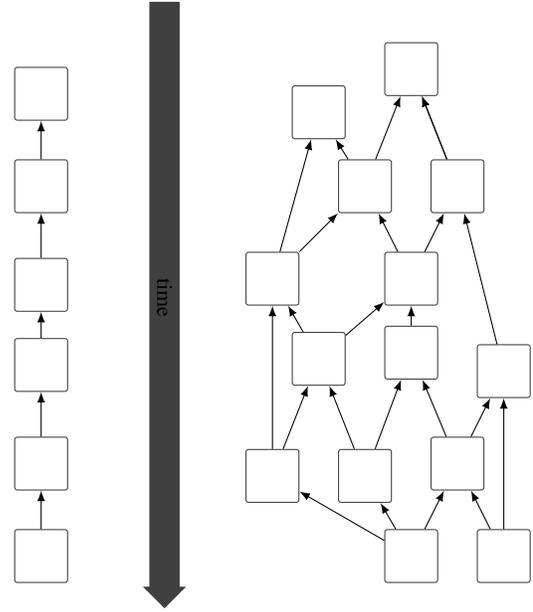
\begin{figure}[t]
    \centering
\begin{tikzpicture}[scale=0.82]
\def\xCoordinate{3.0}
\def\yCoordinate{0.0}
\def\xCoordinateChain{-3.0}
\def\yCoordinateChain{4.5}
\def\yAdd{1.5}
\def\xAdd{0.75}
\def\innerSepar{-1.5}
\def\lineWidth{1.5}
\def\roundedCorners{1}
\def\opacityInternal{0.5}
\def\minHeight{20}
\def\minWidth{20}
\tikzstyle{block}=[draw, rectangle, minimum height=\minHeight pt, minimum width = \minWidth pt, text centered, rounded corners=\roundedCorners pt, draw=darkgray, font=\large]
\def\arrowStyle{-latex}

\node[block] (block1) at (\xCoordinateChain,\yCoordinateChain) {};
\node[block] (block2) at (\xCoordinateChain,\yCoordinateChain-\yAdd) {};
\node[block] (block3) at (\xCoordinateChain,\yCoordinateChain-2*\yAdd-0.1) {};
\node[block] (block4) at (\xCoordinateChain,\yCoordinateChain-3*\yAdd+0.1) {};
\node[block] (block5) at (\xCoordinateChain,\yCoordinateChain-4*\yAdd) {};
\node[block] (block6) at (\xCoordinateChain,\yCoordinateChain-5*\yAdd) {};
\draw[\arrowStyle] (block2) -- (block1);  
\draw[\arrowStyle] (block3) -- (block2); 
\draw[\arrowStyle] (block4) -- (block3); 
\draw[\arrowStyle] (block5) -- (block4); 
\draw[\arrowStyle] (block6) -- (block5); 

\draw[\arrowStyle,path fading=north] (block1) -- (\xCoordinateChain,\yCoordinateChain+\yAdd);
\draw[\arrowStyle, path fading = south]  (\xCoordinateChain,\yCoordinateChain-6*\yAdd) -- (block6) ; 

\node[block] (central_block) at (\xCoordinate,\yCoordinate+0.3) {};
         
\node[block] (blue_below_1_1) at (\xCoordinate-\xAdd,\yCoordinate-\yAdd-0.2) {};
  \draw[\arrowStyle] (blue_below_1_1) -- (central_block);  

\node[block] (blue_below_1_2) at (\xCoordinate+\xAdd,\yCoordinate-\yAdd) {};
  \draw[\arrowStyle] (blue_below_1_2) -- (central_block);

 
 \node[block] (blue_below_2_1) at (\xCoordinate,\yCoordinate-2*\yAdd) {};
  \draw[\arrowStyle] (blue_below_2_1) -- (blue_below_1_2); 
 \draw[\arrowStyle] (blue_below_2_1) -- (blue_below_1_1); 
 
 \node[block] (blue_below_2_2) at (\xCoordinate+2*\xAdd,\yCoordinate-2*\yAdd) {};
 
 \draw[\arrowStyle] (blue_below_2_2) -- (blue_below_1_2); 
 \draw[\arrowStyle, path fading = south]  (\xCoordinate,\yCoordinate-3*\yAdd) -- (blue_below_2_2) ; 
 \draw[\arrowStyle, path fading = south]  (\xCoordinate-3*\xAdd,\yCoordinate-3*\yAdd) -- (blue_below_2_1) ; 
 
 \draw[\arrowStyle, path fading = south]  (\xCoordinate+3*\xAdd,\yCoordinate-3*\yAdd) -- (blue_below_2_2) ;

 \node[block] (red_above_1) at (\xCoordinate,\yCoordinate+\yAdd) {};
  \draw[\arrowStyle] (central_block) -- (red_above_1);

  \node[block] (red_above_2_1) at (\xCoordinate+\xAdd,\yCoordinate+2*\yAdd) {};
  \draw[\arrowStyle] (red_above_1) -- (red_above_2_1);

  \node[block] (red_above_2_2) at (\xCoordinate-\xAdd,\yCoordinate+2*\yAdd) {};
  
  \draw[\arrowStyle] (red_above_1) -- (red_above_2_2);

  \node[block] (red_above_3) at (\xCoordinate,\yCoordinate+3*\yAdd+0.4) {};
     \draw[\arrowStyle] (red_above_2_2) -- (red_above_3);
  \draw[\arrowStyle] (red_above_2_1) -- (red_above_3);

 \node[block] (red_above_3_0) at (\xCoordinate-2*\xAdd,\yCoordinate+2.8*\yAdd) {};
    \draw[\arrowStyle] (red_above_2_2) -- (red_above_3_0);
  \draw[\arrowStyle] (red_above_2_1) -- (red_above_3);
 \draw[\arrowStyle,path fading=north] (red_above_3) -- (\xCoordinate,\yCoordinate+4*\yAdd);
  \draw[\arrowStyle,path fading=north] (red_above_3_0) -- (\xCoordinate-\xAdd,\yCoordinate+4*\yAdd);
  \draw[\arrowStyle,path fading=north] (red_above_3_0) -- (\xCoordinate-2*\xAdd,\yCoordinate+4*\yAdd);

\node[block] (central_grey_left) at (\xCoordinate-2*\xAdd,\yCoordinate+0.2) {};
 \draw[\arrowStyle] (central_grey_left) -- (red_above_1);

\node[block] (central_grey_right) at (\xCoordinate+2*\xAdd,\yCoordinate) {}; 

 \draw[\arrowStyle] (central_grey_right) -- (red_above_2_1);
 \draw[\arrowStyle] (blue_below_1_2) -- (central_grey_right);

 \node[block] (above_grey_1) at (\xCoordinate-3*\xAdd,\yCoordinate+\yAdd) {};
 \draw[\arrowStyle] (above_grey_1) -- (red_above_2_2);
\draw[\arrowStyle] (central_grey_left) -- (above_grey_1);
\draw[\arrowStyle] (blue_below_1_1) -- (central_grey_left);
\draw[\arrowStyle] (above_grey_1) -- (red_above_3_0);

 \node[block] (below_grey_1) at (\xCoordinate-3*\xAdd,\yCoordinate-\yAdd-0.2) {};
\draw[\arrowStyle] (below_grey_1) -- (above_grey_1);
\draw[\arrowStyle] (below_grey_1) -- (central_grey_left);
\draw[\arrowStyle] (blue_below_2_1) -- (below_grey_1);
\draw[\arrowStyle] (blue_below_2_2) -- (central_grey_right);

  \def\xCoordinate{-4.5}
  \def\yCoordinate{\yAdd * 2/3}
  \def\x{3.0}
  \def\y{5.0}
  \def\R{\x+0.004}
  \def\yc{\y+0.02}
  \def\e{0.4}
  \def\opacityCones{0.2}


\node [single arrow,top  color=darkgray, bottom color=darkgray,
single arrow head extend=3pt,transform shape, opacity = 2*\opacityCones, minimum height=280pt, text opacity=1, rotate = 270, anchor=west] 
at (\xCoordinate+\x*7/6,\yCoordinate+\y){time
};
 
\end{tikzpicture}
\caption{Blockchain and blockDAG}
    \label{fig:chainDAG}
    \vspace{-0.3cm}
\end{figure}

\subsection{Why a DAG-based Consensus Protocol}
In the pursuit of transcending the inherent trade-off between security and performance and in response to the performance bottlenecks, DAG-based consensus protocols were proposed as a solution. These protocols promise high scalability and fast confirmation of transactions, hence effectively addressing the intricate balance between security and performance. 
Following, we describe the promises and challenges of DAG-based consensus protocols compared to linear chain protocols.

\textit{Advantages and promises of DAG-based consensus protocols compared to blockchains}
\begin{enumerate}[leftmargin=*]
        \item \textit{Scalability:} DAG-based protocols can process transactions in parallel rather than sequentially, as in blockchains. This capability offers improved scalability and throughput compared to sequential blockchains.
        \item \textit{Latency:} 
        Latency in DAG-based protocols, particularly regarding transaction confirmation time, varies with the underlying consensus mechanism. In PoW-based protocols, a DAG structure enables shorter block times and faster transaction confirmation.
        In some Proof of Stake (PoS)-based protocols, using additional reliable broadcast primitives increases the latency. Therefore, the extent to which DAG-based protocols reduce latency is conditional on the specific consensus approach and the architectural decisions around synchrony and security.
        
        \item \textit{Flexibility:} The DAG architecture allows for more flexible consensus mechanisms and can adapt to various network conditions, potentially making it more versatile than a blockchain. For example, in traditional blockchains, every block serves three roles: acting as a leader that validates transactions, providing content in the form of transaction data, and voting on the causal history.  However, blocks in a DAG structure can have differentiated roles, e.g., see Section \ref{sec:decoupling}, enabling a more distributed approach to consensus and transaction validation.

         \item \textit{Parallel writing:} DAG-based protocols enable concurrent block production, diverging from the leader-based block generation model of traditional blockchains. This architecture permits multiple participants to simultaneously append transactions or blocks to the ledger, effectively broadening writing access. In its most extreme situation, every participant may independently produce blocks, eliminating reliance on a single or group of block producers. This property may remove the need for additional mempools of transactions, reduce the latency, and remove bottlenecks associated with sequential block generation.
    \end{enumerate}

\textit{Challenges in DAG-based consensus protocols compared to blockchains}

\begin{enumerate}[leftmargin=*]
        \item \textit{Security risks:} The increased complexity of DAG structures introduces additional potential attack vectors, as more intricate systems often expose a broader range of vulnerabilities compared to simpler protocols.
        \item \textit{Understanding and adoption:} The more complex nature of DAG-based protocols can make them harder to understand and adopt, particularly for developers accustomed to traditional blockchain technology. Protocols that only provide partial ordering may encounter more difficulties adapting to existing infrastructures.
        \item \textit{Consensus mechanism maturity:} Consensus mechanisms used within DAG-based DLTs are often newer and less tested than those used with blockchains, creating uncertainty around their long-term stability and resilience against evolving network threats.
        \item \textit{Reachability challenge:} The concept of reachability within a DAG refers to the ability of a new block to reference earlier transactions or blocks, which is central to the structure's integrity and efficient functioning. This ability has ramifications for functionalities, such as pruning (removing old data to save space) and the operation of light clients (nodes that do not store the full ledger data).
    \end{enumerate}
The number of DAG-based consensus protocols has been increasing. However, there have been few attempts to consolidate these protocols and provide a comprehensive analysis of them. As the DAG-based DLT community grows, it is becoming increasingly important to have a holistic understanding of the various architectures and trade-offs involved.

\subsection{Contribution}

This SoK offers:
\begin{itemize}[leftmargin=*]
    \item A structured classification of DAG-based consensus protocols into Availability-focused and Consistency-focused categories, analyzing their attributes, trade-offs, and consensus approaches (Sec.~\ref{consensus-protocols}).
    \item A systematic overview of attack vectors relevant to each protocol category, detailing potential vulnerabilities and countermeasures (Sec.~\ref{security-attack}).
    \item An examination of desirable properties currently emphasized in DAG-based DLT research (Sec.~\ref{desirable-properties}).
    \item A discussion of recent advancements in DAG-based DLTs that relate to consensus mechanisms but extend beyond the primary classification (Sec.~\ref{discussion}).
    \item An outline of research gaps and future directions to inform and guide ongoing work in the field (Sec.~\ref{futureResearch}).
\end{itemize}

\textit{Note}: This paper focuses on the conceptual/architectural design of the DAG-based consensus protocols. However, the paper does not discuss or present any performance evaluation of the included protocols.

\subsection{Related Work}
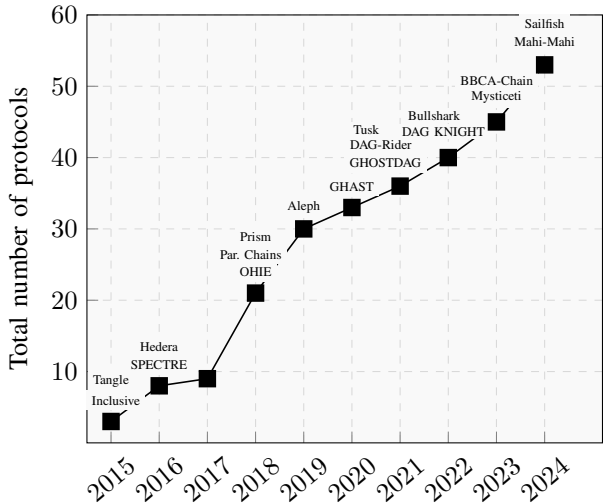
\begin{figure}[th] 
\label{fig:evolutionDAG}
\centering 
\begin{tikzpicture}
\begin{axis}[
ylabel={Total number of protocols},
xmin=2014.5, xmax=2025.2,
ymin=0, ymax=60,
xtick=data,
xticklabel style={/pgf/number format/1000 sep=,rotate=40},
ytick={10,20,30,40,50, 60},
grid=major, 
grid style={dashed,gray!30}, 
axis background/.style={fill=gray!5},
legend style={fill=gray!5, at={(0.5,-0.15)},anchor=north},
xtick pos=lower, ytick pos=left,
]
    \addplot [semithick, mark=square*, black, mark options={scale=1.5}] coordinates {
(2015,3)
(2016,8)
(2017,9)
(2018,21)
(2019,30)
(2020,33)
(2021,36)
(2022,40)
(2023,45)
(2024, 53)
    };

 \node [above, font=\tiny, fill=gray!5, inner sep=2pt] at (axis cs: 2015,7) {Tangle};
    \node [above, font=\tiny, fill=gray!5, inner sep=2pt] at (axis cs: 2015.1,4.5) {Inclusive};
        \node [above, font=\tiny, fill=gray!5, inner sep=2pt] at (axis cs: 2016,12) {Hedera};
         \node [above, font=\tiny, fill=gray!5, inner sep=2pt] at (axis cs: 2016,9.5) {SPECTRE};
            \node [above, font=\tiny, fill=gray!5, inner sep=2pt] at (axis cs: 2018,22.5) {OHIE};
            \node [above, font=\tiny, fill=gray!5, inner sep=2pt] at (axis cs: 2017.9,25) {Par. Chains};
            \node [above, font=\tiny, fill=gray!5, inner sep=2pt] at (axis cs: 2018,27.5) {Prism};
             \node [above, font=\tiny, fill=gray!5, inner sep=2pt] at (axis cs: 2019,31.5) {Aleph};
              \node [above, font=\tiny, fill=gray!5, inner sep=2pt] at (axis cs: 2020,34.5) {GHAST};
    \node [above, font=\tiny, fill=gray!5, inner sep=1pt, yshift=5pt] at (axis cs: 2020.3,41) {Tusk};
    \node [above, font=\tiny, fill=gray!5, inner sep=2pt, yshift=5pt] at (axis cs: 2020.6,38.5) {DAG-Rider};
    \node [above, font=\tiny, fill=gray!5, inner sep=2pt, yshift=5pt] at (axis cs: 2020.7,36) {GHOSTDAG};
      \node [above, font=\tiny, fill=gray!5, inner sep=1pt, yshift=5pt] at (axis cs: 2021.9,40.7) {DAG KNIGHT};
      \node [above, font=\tiny, fill=gray!5, inner sep=1pt, yshift=5pt] at (axis cs: 2021.7,43) {Bullshark};
   \node [above, font=\tiny, fill=gray!5, inner sep=2pt, yshift=5pt] at (axis cs: 2023,45) {Mysticeti};
      \node [above, font=\tiny, fill=gray!5, inner sep=2pt, yshift=5pt] at (axis cs: 2023,47.5) {BBCA-Chain};
      \node [above, font=\tiny, fill=gray!5, inner sep=2pt, yshift=5pt] at (axis cs: 2024,53.0) {Mahi-Mahi};
      \node [above, font=\tiny, fill=gray!5, inner sep=2pt, yshift=5pt] at (axis cs: 2024,55.5) {Sailfish};
   
\end{axis}
\end{tikzpicture}
\caption{The evolution of DAG-based consensus protocols over time.\\ {
   Our estimation of the total number of protocols is based on a manual review of papers citing previous work starting with Hashgraph~\cite{baird2020hashgraph} and Tangle \cite{popov2018tangle}—in Google Scholar and their related work sections. This count, intended as a rough approximation, highlights the growing diversity and interest in DAG protocols.}}
\label{Fig:Evolution}
\end{figure}

Recently, there has been a growing interest in DAG-based protocols in both industrial and academic circles. Numerous DAG-based consensus protocols have been developed to optimize the complex interplay between security and performance. This paper seeks to systematize a wide variety of research on these DAG-based consensus protocols. An overview of the development of these protocols is depicted in Figure~\ref{Fig:Evolution}.

Wang et al.~\cite{wang2023sok} provides a review of DAG-based systems, exploring essential aspects such as consensus mechanisms, security, and performance. However, the study does not address emerging insights and critical open research questions in DAG-based consensus. Furthermore, recent advancements in consistency-focused DAG protocols necessitate an updated, synthesized reference for researchers.

In contrast, \cite{SurveyOnDagLu} organizes DAG-based ledgers into structural categories—main chain, parallel chains, natural topology, and layered DAGs—providing a taxonomy based on the consensus structure. Our SoK introduces a different framework by classifying DAG protocols through an availability and consistency lens. This approach enables a more conceptual understanding of the trade-offs and challenges relevant to DAG-based consensus protocols.

Other studies have examined distributed ledger technologies (DLTs) across a broader spectrum. Bellaj et al.\cite{bellaj2022sok} survey DLTs using a layered model, categorizing them into chained, chainless, and hybrid types, providing a high-level overview that includes both blockchain and non-blockchain structures. Kannengießer et al.\cite{kannengiesser2019mind} analyze various DLT characteristics and the inherent trade-offs between them, highlighting the diverse operational and structural choices within DLT systems. Wu et al.~\cite{wu2022chain} review both chain-based and DAG-based ledgers, offering a taxonomy that differentiates consensus mechanisms and structural variations. They also identify open research questions relevant to each category, giving insights into potential future developments.

\section{DAG-based Consensus in a Nutshell} \label{consensus-nutshell}
A DAG-based consensus protocol arranges blocks in a DAG instead of a linear chain. The directed edges in the DAG establish a causal order linked by cryptographic means, providing evidence of node communication. Nakamoto's proposal already presents this concept of causal dependency, but DAG-based consensus protocols use it more explicitly. 
Crucially, the architecture allows each block to reference multiple predecessors, thereby facilitating the inclusion of more blocks. This contrasts with the traditional blockchain approach, which privileges the blocks along the longest chain.

More precisely, the DAG consists of vertices and directed edges between them. Each vertex represents a block, while the directed edges represent the relationships between these vertices. Specifically, an edge between two vertices represents a partial order relationship, where one vertex verifies, confirms, or witnesses the other vertex. An edge represents a hash reference from one vertex to another. In this way, a reference gives a causal order of the vertices (blocks). The fact that the DAG is acyclic and thus does not contain any cycle guarantees the absence of circular dependencies in our DAG structure. We also refer to Figure~\ref{fig:chainDAG} for an illustrative comparison between a blockchain and a (block)DAG.

In a traditional linear blockchain, the longest chain rule serves two purposes: 1) agreeing on the included blocks and 2) establishing their order.
Transitioning to DAG-based structures necessitates redefining these mechanisms due to the non-linear nature of DAGs. Two principal categories of protocols have emerged: those that perform both tasks directly on the DAG itself and Byzantine Fault Tolerant (BFT) protocols that utilize supplementary broadcast primitives for the agreement on the included blocks. 

Within this SoK, the analysis of principal design decisions is essential. These decisions determine the structural and operational attributes of the underlying DLT systems and impact scalability, security, and efficiency. Such scrutiny is crucial for comparing the attributes and constraints of different DLT architectures. Therefore, we will present these design choices for DAG-based DLTs.

\subsection{Ordering}
In distributed ledgers, consensus is often associated with establishing a total ordering of blocks. However, total ordering is not always essential, particularly for payment systems (e.g., \cite{guerraoui2019consensus, baudet2020fastpay, lewis2023flash}). In these cases, a partial ordering—provided naturally by the underlying DAG structure—can be sufficient to confirm transactions. This approach can significantly improve throughput and efficiency for applications that do not rely on strict sequential processing of transactions.

\subsection{Ledger Model}\label{sec:ledgerModel}
Most consensus protocols are agnostic to the underlying ledger models as they rely on a total ordering of transactions. However, some protocols, like those in SPECTRE~\cite{sompolinsky2016spectre}, IOTA~\cite{muller2022tangle}, and Avalanche~\cite{rocket2019scalable}, deviate by adopting a partial order, challenging the idea that total ordering is indispensable. 

Two prevalent ledger models serve as the backbone for transaction management:
\begin{enumerate}[leftmargin=*]
    \item \textit{UTXO (Unspent Transaction Output):} In a UTXO model, transactions are transfers of value from previous transaction outputs to new unspent outputs in an inductive way. New unspent transaction outputs are called UTXO and are used as inputs for new transactions. Bitcoin~\cite{nakamoto2008bitcoin}, and Cardano~\cite{cardanoEUTXO}, epitomises this approach. It was adapted to the DAG-setting by IOTA~\cite{muller2022tangle} and Avalanche~\cite{rocket2019scalable}\footnote{{The Avalanche crypto project is no longer pursuing the case of UTXOs and their Avalanche consensus protocol on its main net and currently uses an account-based ledger state with a consensus protocol named Snowman.}} since it sidesteps the need for total ordering by structuring the UTXOs as a DAG, promising higher parallelism in transaction processing.

    \item \textit{Account-based:} In an account-based model, each public address is considered an account. Each account has a balance associated with it. Transactions signify direct value transfers between these accounts, enabling a straightforward balance update mechanism exemplified by Ethereum~\cite{wood2014ethereum}.
\end{enumerate}

Object-based and message-based models offer a more general description of possible ledger models. In object-based ledgers, such as the UTXO model, transactions modify objects and thus only affect local states. Due to its localized change impact, this model naturally lends itself to sharding solutions. The causality in object-based models forms a DAG, suggesting that many scenarios do not require a total ordering of transactions thanks to their innate parallelism, as explored in the reality-based ledger model \cite{reality-based-ledger}. Conversely, message-based ledgers conceptualize transactions as messages that induce changes across a global state, highlighting a fundamental distinction in how transactions are processed and effectuated.

A mention of owned and shared objects and how they interplay in contemporary platforms like SUI~\cite{sui_2022_wp} further refines this classification, showcasing a transition towards more granular and flexible state management within DLTs. Such differentiation could enhance understanding and stimulate the development of more scalable ledger solutions, as done in \cite{babel2023mysticeti}. 

\subsection{Consensus Participation and Writing Access}\label{sec:consensus_participation}

The model of consensus participation determines who can validate and contribute to the DAG. This model directly impacts the system's openness, security, and decentralization. Consensus participation generally falls into four categories:

\begin{enumerate}[leftmargin=*]
    \item \textit{Lottery-based:} Participation is determined by a Proof-of-Work (PoW) or Proof-of-Stake (PoS) lottery mechanism, where nodes are selected based on computational work or stake holdings.
    
    \item \textit{Permissioned:} Participation is restricted to a predefined set of validators granted access by the network initiator or governed by protocol-defined rules. All participants are typically aware of one another.
    
    \item \textit{Committee-based:} A rotating subset of validators is selected to participate in consensus for a designated period, allowing for flexibility while maintaining control over participation.
    
    \item \textit{Open Participation:} Any node can join as a validator without permission, allowing fully open participation in the consensus process. 
\end{enumerate}
Recent work by Lewis-Pye and Roughgarden~\cite{lewis2023permissionless} outlines a hierarchy of ``degrees of permissionlessness'' describing how participants' knowledge affects the consensus mechanism. 

While many permissioned protocols could theoretically incorporate committee-based selection, we categorize them as permissioned unless committee selection is explicitly addressed in their theoretical design.

Writing access, or the ability to propose new blocks, typically aligns with consensus participation. However, specific protocols like \cite{muller2022tangle} and \cite{obelia} may allow broader access to block proposals while limiting consensus participation to a subset of nodes, balancing decentralization and efficiency.

\subsection{Network Model}\label{sec:networkModel}
Consensus protocols are designed based on different network models that describe how communication occurs between the participants. These models are often defined by the potential power of an adversary to control message delays. Three commonly used communication models are synchronous, asynchronous, and partially synchronous. To comprehend these models, it is essential to have a formal understanding of the following key concepts in network systems.

The \textit{delay parameter $\Delta$} defines the maximum delay (in time steps) a message suffers in a synchronous phase. The parameter $\Delta$ is known upfront to the nodes in the protocol. The \textit{Global Stabilization Time } (GST) dictates when the underlying communication network switches from asynchronous to synchronous. Unlike $\Delta$, GST is not known upfront in the protocol and can be chosen by an adversary. More precisely:
\begin{enumerate}[leftmargin=*]
    \item \textit{Synchronous:} In a synchronous model, the messages between nodes are delivered within known and predictable time frames. Nodes follow a common global clock or synchronized time intervals to operate in lockstep. In short, for any message sent, an adversary can delay its delivery by at most $\Delta$. This model is used in time-sensitive systems, making it suitable for some DLT-based real-time systems.
    \item \textit{Asynchronous:} In an asynchronous model, the messages between nodes experience varying delays. An adversary can delay the delivery of any message sent by any finite amount of time, but eventually, the message gets delivered. This offers the most flexibility, as message delivery times are unpredictable. This model is standard and used in scenarios where global synchronization is impractical.
    \item \textit{Partially Synchronous:} In a partially synchronous model, messages are delivered within an unknown finite time. In the equivalent \textit{eventually synchrony} model, the GST event will occur after a certain, unspecified time has passed. Moreover, any message sent at time $t$ must be delivered by time $\Delta + max(t, \mathrm{GST})$. This model offers a balance between flexibility and accuracy. Nodes in this model aim to operate with timers that can measure time $\Delta$ after an event, and there is no guarantee about the exact timing of message deliveries. These models are common in network environments where transient conditions may cause disruptions. After GST, the network guarantees that messages are delivered within the delay parameter $\Delta$, providing a window of predictability that can be very useful for consensus algorithms, especially in fault-tolerant systems that must operate under the assumption of periodic network instability. This model is particularly relevant for DLT systems that aim to achieve consensus despite unpredictable network conditions.
\end{enumerate}

\subsection{Dynamic Availability}\label{sec:dynamicAvailability}

In certain consensus protocols, as highlighted by Neu et al. (2022)~\cite{neu2022availability}, nodes operate within a fluid participation framework that aligns well with consensus algorithms built on the ``sleepy'' model, put forth by Pass and Shi (2017)~\cite{pass2017sleepy}. In these consensus protocols, nodes can exhibit dynamic availability, seamlessly transitioning between periods of activity and dormancy up to a specified \textit{Global Awake Time} (GAT). After this moment, an assumption is made that all honest nodes will be consistently online, facilitating consensus in an environment reflective of real-world participation and connectivity patterns. 
Thus, dynamic availability ensures transactions are processed and finalized despite any transient fault.

\subsection{Additional Broadcast Primitive}\label{sec:RB}
 DAG-based consensus protocols adopt different strategies to ensure the reliable propagation of blocks. These strategies can be broadly classified based on their reliance on additional reliable broadcast primitives.

Many DAG-based systems use established Reliable Broadcast (RB) protocols~\cite{bracha1987asynchronous} to distribute blocks. These systems implement an RB protocol as an additional layer, ensuring consistent and reliably disseminated information among nodes, even in the face of network challenges or adversarial behaviour.

Conversely, some protocols employ the blockDAG structure itself to replicate the functionality of a reliable broadcast mechanism. The DAG's architecture inherently distributes data across the network so that blocks are organically verified and propagated without needing a specialized RB protocol.

Both approaches aim to maintain a secure and coherent state across distributed participants. However, they differ fundamentally in whether they view the blockDAG as sufficient for reliable broadcasting or require an additional overlay of RB protocol to strengthen information dissemination and security. The additional use of RB typically influences latency by increasing it for several network rounds.

\subsection{Adversarial Model}\label{sec:adversarialModel}

The adversary model defines attacker capabilities essential for setting security parameters and building resilient consensus mechanisms. Typically, these models assume adversaries with bounded computational power, limited to polynomial-time algorithms or unbounded power, capable of breaking cryptographic primitives. 
This model is intrinsically connected to the network's communication model. The FLP impossibility result, \cite{FLP}, highlights the challenges of achieving consensus in a purely asynchronous network with just one faulty node. To address this, DLTs may assume partial synchrony or incorporate elements of randomness to secure consensus against adversarial actions within probabilistic bounds.
In systems that use PoW and PoS offering probabilistic finality, the critical threshold is $50\%$. This means that if more than $50\%$ of the participants are honest and not faulty, the network is operational, guaranteeing safety and liveness. On the other hand, BFT mechanisms adopt a $2/3$ threshold, requiring more than $2/3$ of honest and non-faulty nodes, and offer a deterministic finality. These two thresholds are fundamental to the security models of DLTs and reflect a balance between resilience to adversarial control and consensus outcome under varying assumptions of network synchrony.

\subsection{Leader-based V/s Leaderless Consensus}\label{sec:leader}
Consensus mechanisms within DLTs typically adopt either a leader-based or a leaderless structure. Leader-based consensus protocols designate a leader to propose an orderly view of the ledger transactions. This view becomes committed when a sufficient consensus, or votes, affirming the view is reached. Leaders may be chosen through various mechanisms, such as PoW/PoS lottery or a round-robin selection.

In contrast, leaderless consensus protocols decentralize the proposal process, with decisions emerging from the collective input of all nodes.

\begin{table*}[!t]
\centering
\Huge
\caption{Overview of DAG-based Consensus Protocols}
\label{tab:dag_comparison_mentioned_transposed}
\resizebox{\textwidth}{!}{%
\begin{tabular}{|p{8.5cm}|p{4.5cm}|c|c|c|c|c|c|l|l|p{4.0cm}|}
\hline
\textbf{Protocol} & \textbf{Participation} & 
\textbf{\makecell{Additional \\ Broadcast \\Primitive}} & \textbf{\makecell{Additional \\ Mempool}} & \textbf{\makecell{Leader-\\based}} & \textbf{\makecell{Dynamic \\ Availability}} & \textbf{Ordering} & \textbf{Finality}\footnotemark & \textbf{Network Model}\footnotemark  & \textbf{Innovation/Relation} \\ \hline
\hline
Nakamoto~\cite{nakamoto2008bitcoin} & PoW & No & Yes & Yes & Yes & Total & Probabilistic & Synchronous & Genesis of blockchain and cryptocurrencies \\[0.25cm] \hline

GHOST~\cite{sompolinsky2015secure} & PoW & No & Yes & Yes & Yes & Total & Probabilistic & Synchronous\footnotemark[3] & Improved upon Nakamotos's chain structure for higher throughput using heaviest chain selection \\[0.25cm] \hline

Inclusive~\cite{lewenberg2015inclusive} & PoW & No & Yes & Yes & Yes & Total & Probabilistic & Synchronous\footnotemark[3] & Evolved from GHOST, presents a game-theoretic model for reward among miners. \\[0.25cm] \hline 

Byteball~\cite{churyumov2016byteball} & Open & No & Yes & No & Yes & Total & Deterministic & Synchronous\footnotemark[3] & Witness nodes for syncing and ordering transactions using main chain index (MCI) \\[0.25cm] \hline

SPECTRE~\cite{sompolinsky2016spectre} & PoW & No & Yes & No & Yes & Partial & Probabilistic & Partial Synchronous & Further evolved from GHOST, prioritizing fast confirmations using pairwise voting \\[0.25cm] \hline

PHANTOM~\cite{sompolinsky2021phantom} & PoW & No & Yes & No & Yes & Total & Probabilistic & Synchronous & Introduced k-cluster for better structure over SPECTRE's DAG \\[0.25cm] \hline

GHOSTDAG~\cite{sompolinsky2021phantom} & PoW & No & Yes & No & Yes & Total & Probabilistic & Synchronous & Focused on reducing latency and refined k-cluster approach \\[0.25cm] \hline

DAG KNIGHT~\cite{sompolinsky2022dag} & PoW & No & Yes & No & Yes & Total & Probabilistic & Partial Synchronous\footnotemark[3] & Parameterless expansion from GHOSTDAG without latency bounding \\[0.25cm] \hline

Prism~\cite{Prism} & PoW & No & Yes & Yes & Yes & Total & Probabilistic & Synchronous & Separated consensus functions into multiple parallel chains \\ \hline

GHAST~\cite{GHAST} & PoW & No & Yes & Yes & Yes & Total & Probabilistic & Synchronous & Tree-graph approach and adaptive mechanisms for performance and security \\[0.25cm] \hline

Hashgraph~\cite{baird2020hashgraph} & Permissioned & No & No & No & No & Total & Deterministic & Asynchronous & First optimistic DAG with fair ordering derived from block's timestamps  \\[0.25cm] \hline

Jointgraph~\cite{xiang2021jointgraph} & Permissioned & No & No & Yes & No & Partial & Deterministic &  Synchronous\footnotemark[3] & Improves throughput and latency of Hashgraph using a supervisor node with additional events \\[0.25cm]  \hline

Aleph~\cite{gkagol2019aleph} & Permissioned & Yes & Yes & No & No & Total & Deterministic & Asynchronous &  Round-based DAG with retrospective randomized leader selection \\[0.25cm]  \hline

DAG-Rider~\cite{keidar2021all} & Permissioned   & Yes & Yes & No & No & Total & Deterministic & Asynchronous & Reduced latency compared to Aleph by a new commit rule \\[0.25cm]  \hline

Avalanche~\cite{rocket2019scalable} & Permissioned  & No & No & No & No & Partial & Probabilistic & Synchronous & Introduced leaderless consensus using repeated sub-sampled querying \\[0.25cm]  \hline

OHIE~\cite{yu2020ohie} & PoW & No & Yes & No & Yes & Total & Probabilistic & Synchronous & Many parallel instances of the
Nakamoto consensus 
\\[0.25cm] \hline
Parallel Chains~\cite{fitzi2018parallel}
& PoW / PoS & No & Yes & No & Yes & Total & Probabilistic & Partial synchronous & theoretical framework for optimistic throughput
\\[0.25cm] \hline

Expected consensus~\cite{wang2023security} & PoS & No & Yes & No & Yes & Total & Probabilistic & Synchronous  & Increased throughput and adapted consensus for storage resource\\[0.25cm] \hline

Nano~\cite{lemahieu2018nano} & Open & No & Yes & No & Yes & Partial & Deterministic & Synchronous\footnotemark[3]  & Block-lattice structure allowing asynchronous updates \\[0.25cm] \hline

Chainweb~\cite{martino2018chainweb} & PoW & No & Yes & No & No & Partial & Probabilistic & Synchronous\footnotemark[3]  & Employs simple payment verification (SPV) for token transfer \\[0.25cm] \hline

Meshcash~\cite{bentov2021tortoise} & PoW & No & Yes & No & Yes & Total & Deterministic & Asynchronous & Ebb-and-flow type protocol \newline \\[0.25cm] \hline

Tusk~\cite{danezis2022narwhal} & Permissioned & Yes & Yes & No & No & Total & Deterministic & Asynchronous & Mempool management by Narwhal, resulting in high throughput; Tusk orders digests of blocks \\[0.25cm] \hline

Bullshark~\cite{spiegelman2022bullshark} & Permissioned & Yes & Yes & Both & No & Total & Deterministic & Partially Synchronous / Asynchronous & Improved latency by introducing new commit rule; garbage collection and timely fairness after GST \\[0.25cm] \hline

Cordial Miner~\cite{keidar2022cordial} & Permissioned & No & No & Both & No & Total & Deterministic & Asynchronous & Reduced latency by using best effort broadcast and committing leader blocks \\[0.25cm] \hline

Tangle 2.0~\cite{muller2022tangle} & PoS & No & No & No & No & Partial & Deterministic & Synchronous\footnotemark[3] & Conflict resolution using the DAG and reality-based ledger state \newline \\[0.25cm] \hline

BBCA-Chain~\cite{malkhi2023bbca} & Permissioned & Yes & Yes & Yes & No & Total & Deterministic & Partially Synchronous & Latency reduction by several network trips by using a new BBCA primitive for leader blocks \\[0.25cm] \hline

Mysticeti~\cite{babel2023mysticeti} & Permissioned & No & No & Yes & No & Total & Deterministic & Partially Synchronous & Pipelined leader schedule, fast finality for owned object transactions, epoch-closing mechanism \\[0.25cm] \hline

Shoal~\cite{spiegelman2023shoal} & Permissioned & No & No & Yes & No & Total & Deterministic & Partially Synchronous & Improved the latency compared to Bullshark by pipelining leader blocks \\[0.25cm] \hline

Shaol++~\cite{arun2024shoal++} & Permissioned & No & No & Yes & No & Total & Deterministic & Partially Synchronous & Optimised latency by operating multiple DAGs in parallel. \\[0.25cm] \hline

Sailfish~\cite{shrestha2024sailfish} & Permissioned & No & No & Yes & No & Total & Deterministic & Partially Synchronous & Reduced latency compared to Shoal by having multiple leaders every round and novel commit rule \\[0.25cm] \hline

Mahi-Mahi~\cite{Mahi-Mahi} & Permissioned & No & No & Yes & No & Total & Deterministic & Asynchronous & Reduced latency compared to Cordial miners through boost rounds \\[0.25cm] \hline

Slipstream~\cite{Slipstream} & Permissioned & No & No & No & Yes & Total & Deterministic & Partially Synchronous & Ebb-and-flow DAG-based protocol and fast path confirmation on DAG \newline \\[0.25cm] \hline

\end{tabular}%
}
\end{table*}

\footnotetext[2]{The finality results included here represent claims or descriptions by the respective protocol authors. Many protocols listed lack complete formal proofs within their stated models, and results are often based on suggested behaviour rather than rigorous validation.}

\footnotetext[3]{The network model is not always given explicitly in the corresponding papers. In this case, we attempt to classify it based on the information in the paper. These cases are marked by this footnote. }

\subsection{Mempool}\label{sec:mempool}
For protocols where writing access is probabilistic or requires selection from a group, such as lottery-based or committee-based systems, a mempool is essential.

A mempool, short for memory pool, is a temporary storage for transactions broadcasted to the network but not yet included in a block. The mempool serves several important roles:
\begin{enumerate}
    \item \textit{Transaction Queue:} It acts as a queue for pending transactions, maintaining them in an accessible state until they are selected and confirmed by a node with writing access.
    \item \textit{Prioritization:} The mempool can prioritize transactions based on specific criteria, such as fee rates, ensuring that transactions with higher fees are confirmed more quickly.
    \item \textit{Availability:} It ensures that transactions remain available for selection and confirmation, even when uncertain which node will write the next block into the ledger.
    \item \textit{Synchronization:} In systems with multiple potential writers or validators, the mempool is crucial for synchronizing the state across these entities, allowing them to view and validate the same set of unconfirmed transactions.
\end{enumerate}
Some protocols use a blockDAG as a distributed mempool for pending transactions, eliminating the need for a separate mempool structure. The blockDAG then maintains a real-time ``ledger'' of unconfirmed transactions, making the system more transparent and accessible. This more structured mempool can mitigate transaction duplicates; see also discussion in Section \ref{futureResearch}.

\subsection{Finality}
Finality refers to the point at which a transaction or a block of transactions is considered irreversible, permanently part of the ledger, and its execution is settled. Finality comes in two forms: probabilistic and deterministic (or absolute). 

\begin{enumerate}
    \item \textit{Probabilistic Finality:} 
    It means the order of a block can be reverted, but the risk decreases as it gets more embedded in the blockchain or blockDAG. In PoW systems, finality is probabilistic, and in most cases, the probability of a block being reverted decreases exponentially over time.
    \item \textit{Deterministic Finality:} It ensures that the order of a block or the presence of a transaction in the ledger is permanent once it is recorded. Protocols that use BFT ideas usually offer deterministic finality. 
    They require an agreement from at least a supermajority of the validators, making it irrevocably part of the ledger.
\end{enumerate}
Additionally, some DLTs with DAG structures may initially only offer probabilistic finality. However, they can obtain deterministic finality by designating special roles; for instance, witness nodes in Byteball~\cite{churyumov2016byteball} or nodes creating snapshot chains in Vite~\cite{liu2018vite} and IOTA~2.0~\cite{iota}.

\subsection{Overview}
Table~\ref{tab:dag_comparison_mentioned_transposed} presents a diverse range of DAG-based consensus protocols, their underlying models, technical specifications/features, and key innovation ideas. The column ``Innovation/Relation'' highlights the novel contributions of each protocol or delineates its evolution from preexisting technologies.

\section{DAG-based Consensus protocols} \label{consensus-protocols}
In distributed systems, consensus is crucial for agreeing on a single version of the system's state. In DAG-based DLTs, nodes collaboratively build a blockDAG and agree on a subset of interconnected blocks called a prefix\footnote{A prefix in this context is a set of blocks in a DAG which is closed by causal order relations imposed by the DAG.} to ensure a shared understanding of the ledger's history. Once nodes agree on this prefix, they can establish a total ordering for the included transactions. In certain protocols, they might only ascertain a partial ordering, which can suffice for transaction execution depending on the chosen ledger model, see Section \ref{sec:ledgerModel}.

As we develop our systematization, it's essential to recognize the fundamental limitations imposed by the CAP theorem. This theorem states that any distributed system can achieve a maximum of two out of three key properties - consistency, availability, and partition tolerance - at the same time. Our SoK is based on this principle, which guides our analysis and comprehension of the inherent trade-offs and design choices. It's important to note that these properties are defined originally in ~\cite{brewer2000towards}, but we emphasize the need to interpret them in the current context, e.g.,~\cite{CAP12years}. The exact definition varies, and we give the definition provided by Brewer in \cite{CAP12years}.

\begin{table*}[!htbp]
\centering
\caption{Categorization of Consensus Protocols in DAG-based DLTs}
\label{tab:consensusDAG}

\begin{tabular}{|c|c||c|c|}
\hline
\multicolumn{4}{|c|}{\textbf{DAG-based consensus protocols}} \\
\hline
\multicolumn{2}{|c||}{ Availability-Focused } & \multicolumn{2}{c|}{ Consistency-Focused } \\
\hline
 Structured DAG &  Unstructured DAG &  Optimistic DAG & Certified DAG \\
\hline
Expected consensus~\cite{wang2023security} & SPECTRE~\cite{sompolinsky2016spectre} & Hashgraph~\cite{baird2018hedera, baird2020hashgraph} & Aleph~\cite{gkagol2019aleph} \\
Inclusive~\cite{lewenberg2015inclusive}& PHANTOM~\cite{sompolinsky2018phantom} & Jointgraph~\cite{xiang2021jointgraph} & DAG-Rider~\cite{keidar2021all} \\
GHAST/Conflux~\cite{GHAST, li2020decentralized} & GHOSTDAG~\cite{sompolinsky2021phantom} & Cordial Miners~\cite{keidar2022cordial} & Tusk~\cite{danezis2022narwhal} \\
OHIE~\cite{yu2020ohie} & DAG KNIGHT~\cite{sompolinsky2022dag}& Tangle 2.0~\cite{muller2022tangle} & Bullshark~\cite{spiegelman2022bullshark} \\
Chainweb~\cite{martino2018chainweb}& Slipstream~\cite{Slipstream}& Mysticeti~\cite{babel2023mysticeti} & BBCA-Ledger~\cite{stathakopoulou2023bbca} \\
Parallel chains~\cite{fitzi2018parallel} & Byteball~\cite{churyumov2016byteball} & Slipstream~\cite{Slipstream} & BBCA-Chain~\cite{malkhi2023bbca}\\
Prism~\cite{Prism} & Meshcash~\cite{bentov2021tortoise} & Mahi-Mahi~\cite{Mahi-Mahi}& Shoal, Shoal++~\cite{spiegelman2023shoal, arun2024shoalhighthroughputdag} \\
 & Graphchain~\cite{boyen2018graphchain} & & Sailfish~\cite{Shrestha2024SailfishTI} \\
 & Avalanche~\cite{avalanche_2020} & & \\
\hline
\end{tabular}
\end{table*}

The CAP theorem states that any networked shared-data system can have at most two of three desirable properties:

\begin{itemize} [leftmargin=*]
\item \textit{Consistency (Safety):} consistency (C) equivalent to having a single up-to-date copy of the data;

\item \textit{Availability (Liveness):} high availability (A) of that data (for updates); and

\item \textit{Partition Tolerance:} tolerance to network partitions (P).
\end{itemize}

In the above, we informally correlate ``consistency'' with ``safety'' and ``availability'' with ``liveness,'' despite recognising that these terms traditionally describe distinct concepts within distributed systems literature. This correlation hopefully highlights the most essential trade-offs in DAG-based DLT systems.

Given these constraints, DAG-based consensus protocols face trade-offs and are classified into two primary classes. 

\begin{enumerate} [leftmargin=*]
   \item \textit{Availability-focused Protocols:} \\
    Focusing primarily on maintaining the continuous progress of the ledger, most of these protocols use variations of a weighted\footnote{For instance, the weight could be computing power, stake, or storage.} lottery, where nodes are randomly selected to propose new blocks to the DAG and vote on the existing ones. Such protocols are typically suited for permissionless or dynamically available settings~\cite{lewis2023permissionless}. In this setting, a potentially large number of entities participate in the consensus process, and the list of active participants is unknown and changing over time. Canonical examples of availability-focused consensus protocols among linear blockchains are Nakamoto consensus~\cite{nakamoto2008bitcoin} and Ouroboros~\cite{david2018ouroboros}. When applied in a DAG-based architecture, this approach manifests in different forms, including structured and unstructured DAGs.
    It is worth noting that, if guaranteed, finalization in availability-focused protocols is often probabilistic. 
    \begin{enumerate} [leftmargin=*]
         \item \textit{Structured DAG:} These protocols enforce strict block addition rules, keeping an organized DAG topology.
        \item \textit{Unstructured DAG:} These protocols represent a more flexible approach to DAG construction, where nodes have greater liberty in attaching new blocks without stringent adherence to predefined rules. 
        However, the trade-off often lies in increased complexity for achieving consensus and validating transactions. 
    \end{enumerate}

    \item \textit{Consistency-focused Protocols:} \\
    The protocols of this type prioritize ledger consistency by drawing from classical Byzantine Fault Tolerant (BFT) protocols~\cite{bracha1987asynchronous,castro1999practical, lamport1982byzantine}.
    Consistency is mostly achieved in permissioned and quasi-permissionless environments, e.g., see~\cite{lewis2023permissionless}.
    These protocols can be further categorized based on how participants disseminate their blocks.
    \begin{enumerate}  [leftmargin=*]
    \item \textit{Optimistic DAG:} In these protocols, block dissemination is done through the best-effort broadcast, i.e. sending blocks to other nodes without providing any guarantees about the uniqueness of the blocks. Thereby, such protocols require mechanisms to handle equivocation.
    \item \textit{Certified DAG:} These protocols employ reliable and consistent broadcast primitives for block dissemination. This means that every block in a locally maintained DAG can be seen as a certificate signed by a quorum of nodes. Such protocols ensure participants maintain a consistent DAG view.
    \end{enumerate}
\end{enumerate}

We depict Table~\ref{tab:consensusDAG} to show which consensus protocols fall in which category. The subsequent subsections will explore these categories, analyzing key protocols and their innovations.

\subsection{Availability-focused Protocols with Structured DAG}
\subsubsection{Parallel instances of Nakamoto consensus}
The most straightforward approach to improve Bitcoin's scalability is to execute $m$ concurrent instances of the Nakamoto consensus protocol as proposed in \textit{Parallel Chain}s~\cite{fitzi2018parallel} and \textit{OHIE}~\cite{yu2020ohie}. Mining simultaneously and uniformly for all $m$ chains is achieved by a similar technique in both protocols. A newly mined block must include the digests of the latest blocks from all chains, and a miner remains unaware of which chain a new block will extend until it solves the PoW puzzle. After successful mining, the block's hash value decides which chain it belongs to. 

Establishing total ordering on parallel chain architectures can be achieved in different ways. In OHIE, a ranking system for blocks is used. Each block is equipped with a \textit{current rank} and a \textit{next rank}; the current rank mirrors the next rank of its referenced predecessor, while the next rank represents the highest current rank observed across all chains. Parallel Chains employs a different strategy, where one specific chain is designated for synchronization. Each block in the parallel architecture references a block from this synchronization chain. The chronological order of the entire network is then inferred based on the sequence of blocks in this special chain.

Parallel Chains and OHIE guarantee the same safety properties as the Nakamato Consensus, \cite{nakamoto2008bitcoin}, and improve either the throughput or latency, as their scalability does not offer both desired features simultaneously. Multiple other protocols can also fall into this category, e.g., Chainweb~\cite{martino2018chainweb}, but they degrade in security or latency compared to Nakamoto consensus. Nano~\cite{lemahieu2018nano}, Vite~\cite{liu2018vite} protocols follow a similar concept where participants maintain their parallel chains.

\subsubsection{Decoupling block's functionalities}\label{sec:decoupling}
A conceptual approach suggested in \textit{Prism}~\cite{Prism} to augment scalability in parallel chain architectures involves deconstructing the multifaceted functionalities of blocks, as seen in Nakamoto’s consensus. Traditionally, blocks in Bitcoin simultaneously fulfil three key roles: 1) leader election, 2) transaction proposing, and 3) voting on the causal history through parent links. Prism proposes a clear segregation of these functions into distinct types of blocks, each dedicated to a specific purpose, such as transaction processing, block proposal, and voting.

This segregation results in the formation of different types of chains within the architecture. Primarily, it creates one \textit{proposer} chain, where each block references a preceding block within the proposer chain and several \textit{transaction} blocks. In parallel, $m$ \textit{vote} chains emerge, with each chain functioning to cast votes determining the total order of blocks within the proposer chain. Such a structural reorganization clarifies the roles within the blockchain and allows high throughput, low (constant) latency, and the same safety guarantees as in the Nakamoto consensus.

Note that there are predecessors of Prism where functionalities of blocks are decoupled, e.g., FruitChain~\cite{pass2017fruitchains}, Bitcoin-NG\cite{eyal2016bitcoin}, but have not improved some other aspects, e.g., latency.

\subsubsection{Tree-structure in a DAG}
In \textit{GHAST} (Greedy Heaviest Adaptive SubTree)~\cite{GHAST, li2020decentralized}, each block except genesis has precisely one outgoing parent edge and can have multiple outgoing reference edges. The parent edges construct a tree within the broader DAG, and the reference edges establish temporal precedence, indicating that the referenced block predates the block containing the reference. Once a winning chain within the tree is selected, a deterministic order over all referenced blocks can be established. The selection of this winning chain is influenced by the weights assigned to each block, akin to the GHOST protocol~\cite{sompolinsky2015secure}. Here, the fork choice rule systematically favours the heaviest subtree at every fork, typically equating block weight with the extent of computational work done.

GHAST diverges from traditional protocols in its adaptive response to potential liveness attacks. During such scenarios, it modifies the weight distribution among blocks. While normally all blocks carry equal weight, under attack detection, the protocol randomly selects a few blocks to assign non-zero weights and keeps the expected block weight at the same level.

In \textit{Inclusive}~\cite{lewenberg2015inclusive}, the main idea was to include transactions from all the blocks in the ledger. An inclusive rule is defined to select a main chain from the DAG, and further non-conflicting blocks from the outside DAG are appended to the final block structure.
The Inclusive protocol also rewards miners whose blocks are not in the main chain but are contained within the DAG.

\subsubsection{Round-based chain of tipsets}
Another attempt to improve the latency of Nakamoto's consensus was suggested in \textit{Expected Consensus}~\cite{wang2023security}. This heaviest-chain style protocol operates in rounds. Each round involves a cryptographic sortition on the weighted list of participants. It is parametrized such that, on average, a given number of $m$ participants may be eligible to propose a block every round. Every block must reference a \textit{tipset}, a set of blocks sharing the same round and a parent tipset. Every block adds weight (determined by the sortition) to the chain of tipsets, and the consensus is achieved by following the heaviest chain of tipsets. 

\subsection{Availability-focused Protocols with Unstructured DAG}

\subsubsection{Unstructured Block-DAG protocols}\label{sec: unstructured blockdag} 

The \textit{GHOST} protocol~\cite{sompolinsky2015secure} diverges from Nakamoto's longest chain rule in selecting the Greedy Heaviest-Observed Sub-Tree. This enables shorter intervals between block creations without compromising security, a concept explored in \cite{kiayias2019trees} in further detail. GHOST capitalizes on the proof of work in ``off-chain'' blocks,  traversing the tree-like structure that emerges from chain forks. This alternate selection method for the main chain is specifically tailored to mitigate issues associated with network latency.

The \textit{SPECTRE}~\cite{sompolinsky2016spectre} protocol focuses on building a blockDAG structure with separate mechanisms for mining and consensus, leading to a notion of weak liveness for higher scalability. SPECTRE uses a recursive weighted voting technique where each new block in the DAG submits votes (preference ordering) over every pair of blocks based on which block they believe occurred first. The final ordering in SPECTRE is based on the majority vote on pairwise ordering across all the blocks. However, this ordering is not extendable to a total order due to Condorcet paradox~\cite{gehrlein1983condorcet}. While SPECTRE only achieves partial ordering, it works under a partial synchronous network model. This partial ordering makes SPECTRE suitable for payment systems but not for systems that rely on account-based ledgers. 

\textit{PHANTOM}~\cite{sompolinsky2018phantom} is an extension of the SPECTRE protocol. It aims to achieve a total ordering but under a synchronous network model. PHANTOM works by separating block creation or mining from the consensus mechanisms. It builds on the intuition that blocks created by honest miners are well-connected, while adversarial blocks are not well-connected and thus should be excluded.
The notion of $k$-cluster expresses this well-connectedness. Once this $k$-cluster is determined, a total block order is established via a topological ordering. The synchronicity assumptions become apparent in the choice of the parameter $k$, which depends on the latency. 
Moreover, the identification of such $k$-cluster is NP-hard, GHOSTDAG, \cite{sompolinsky2021phantom}, is proposed as a practical alternative; it also addressed possible attack vectors, pointed out by \cite{li2018scaling}.

\textit{DAG KNIGHT}~\cite{sompolinsky2022dag} is an evolution of GHOSTDAG. Unlike GHOSTDAG, DAG KNIGHT assumes no upper bound on network latency. It leverages a dual min-max optimization approach to dynamically find the largest $k$-cluster for each $k$, selecting the minimal $k$ covering at least $50\%$ of the DAG. This design accommodates latency variations for safety and provides a responsive protocol to actual network latency rather than a hard-coded latency bound as in GHOSTDAG.

\textit{Meshcash}~\cite{bentov2021tortoise} is a framework containing two layered protocols to reach a consensus on the blocks added to the DAG using PoW. The blocks are arranged in layers; each block belongs to a layer and refers to blocks in the previous layer. The consensus protocol is an ebb-and-flow style consensus protocol~\cite{neu2021ebb} with a slower protocol 
that favors safety and a faster one that favors liveness. 
The slower protocol works for the blocks in the far past (in old layers), where a weighted voting is performed to decide on the consistency of the blocks.

Tangle 2.0~\cite{muller2022tangle} employs a unique approach to conflict resolution, preventing the locking of funds by actively voting on conflicts within the DAG. Leveraging the reality-based ledger~\cite{reality-based-ledger}, transactions can be treated optimistically, allowing nodes to build on unresolved or unconfirmed transactions. This approach enables concurrent versions to coexist within the ledger without requiring strict total ordering.

The \textit{Slipstream} protocol,~\cite{Slipstream}, is also ebb-and-flow style consensus protocol~\cite{neu2021ebb}. As such, it offers two types of block orderings: an optimistic ordering, which is live and secure in a sleepy model under up to $50\%$ Byzantine nodes, and a final ordering, which is a prefix of the optimistic ordering and ensures safety and liveness in an eventual lock-step synchronous model under up to $33\%$ Byzantine nodes. Slipstream utilizes wall clocks instead of Lamport clocks, allowing for dynamic availability in the network.
Each node in Slipstream generates commitments based on its view of the local DAG and the online nodes within fixed-duration time intervals (rounds). Each commitment represents a hash of the accepted data (e.g., blocks and transactions) within that round, linking to the commitment of the previous round to form a commitment chain. This chain, representing the prefix of a DAG, allows the network to progress and maintain the available ledger state even during partitions. When a partition resolves and a supermajority of validators is online, nodes merge the commitment chains, resuming finalization.
Slipstream also incorporates a UTXO payment system that enables fast transaction confirmation independently of block ordering. During synchrony, transactions can be confirmed in three rounds, with unconfirmed double spends resolved using the DAG structure in a novel approach. 

\subsubsection{Unstructured Transaction-DAG protocols} This class includes protocols that append transactions directly in their DAGs without the block as a wrapper, e.g., Graphchain~\cite{boyen2018graphchain}.

The \textit{Byteball}~\cite{churyumov2016byteball} protocol allows users to add transaction units to the DAG by signing them.
These units are linked in the DAG, containing hashes of prior units to confirm their validity and create a partial order. Users choose trustworthy nodes, called witnesses, based on their reputation. These witness nodes periodically generate transaction units that help compute a main chain in the DAG. Consensus is then found by the total ordering of the transaction using the main chain.

\textit{Avalanche}~\cite{avalanche_2020} protocol allows nodes to repeatedly query a random group of nodes about the validity of a transaction. Further, to be able to vote on several transactions at the same time, the nodes employ a DAG structure. When a node promptly accumulates sufficient positive responses for its query, it solidifies its decision. Critiques have emerged regarding the security (especially liveness) of the protocol \cite{MuPo:22, amoressesar2022spring}, and the Avalanche project stopped working on the DAG-version of the consensus and maintains a chain version of it.

\subsection{Consistency-focused Protocols with Optimistic DAG}
The protocols in this category rely on validators, a special subset of nodes with a non-zero weight in the voting process. 
Blocks are propagated using best-effort broadcast or gossiping protocol. During this propagation, the nodes need to ensure that the recipient of the new block also knows the causal history of the block. 
In optimistic cases, this can result in fewer network trips to achieve block finality. Typically, these protocols can tolerate adversarial validators holding less than $1/3$ of the total weight. It will be convenient to refer to validators holding more than $2/3$ of the total weight as a \textit{supermajority}, and the main assumption in the following protocols is there exists an honest supermajority of validators.

One main problem of reliable or consistent broadcasting blocks among the validators is the problem of equivocations. This describes the situation where a validator presents different blocks to different parts of the network. In the following protocols, the DAG structure is used to avoid equivocations using a mechanism of double approvals. 
Specifically, a block $X$ is \textit{approved} by a block $Y$ if $X$ is reachable from $Y$ in the DAG  and the block creator of $X$ does not equivocate in the causal history of block $Y.$ A block $X$ is \textit{doubly approved} by a block $Y$ if a supermajority of blocks exists in the causal history of block $Y.$ Each validator can be shown to have at most one block in a given round, which is doubly approved by a supermajority of nodes. The idea was introduced in \textit{Hashgraph}\cite{baird2018hedera, baird2020hashgraph}, and its variation was used in other protocols.

Hashgraph suggests constructing a DAG by gossiping over gossiping events\footnote{In Hashgraph, the term an \textit{event} is used instead of a block}. In this protocol, each node maintains its chain of events. When receiving an event block from another node, it randomly selects at least one peer to disseminate information about its current knowledge about blocks, e.g., a new block within the maintained chain is created which in addition references the received block. A node advances to the next round once its block doubly approves a supermajority of blocks from the previous round.

By employing this idea of double approval in the resulting DAG structure, the Hashgraph protocol decides when the network reliably receives each event. The \textit{received} timestamp is then assigned, and the total ordering is achieved by sorting events over the received timestamps.

\textit{Jointgraph}~\cite{xiang2021jointgraph} reduces the number of voting rounds in Hashgraph by introducing a \textit{supervisor node}. An event in Jointgraph is final if the event receives more than $2/3$ votes from the nodes in which one of the votes is from the supervisor node. The supervisor node also creates snapshots and storage events to finalize the events from the ordinary nodes and provide a total order. When the supervisor node is offline, Jointgraph switches to the Hashgraph process.

\textit{Cordial Miners}~\cite{keidar2022cordial} is a family of simple and efficient DAG-based consensus protocols with instances for the asynchronous and eventual synchronous models.
Similar to Hashgraph, the core idea of Cordial Miners is to utilize the constructed optimistic DAG, called the \textit{blocklace}, for different tasks such as block distribution, equivocation-exclusion, and block ordering. While Hashgraph uses timestamps for ordering, Cordial Miners commit leader blocks, and the slices between committed leader blocks result in ordering.  The blocklace is fragmented into waves using the rounds assigned to blocks. Each wave could have at most one committed leader block chosen using either a round-robin for a partially synchronous network model or a retrospective random selection for an asynchronous network model.

It has been shown~\cite{guerraoui2019consensus} that a system that enables participants to make simple payments needs to solve a simpler task. In cases where payments are independent of one another (e.g. single-owned token assets or UTXO transactions), there is no necessity to order them, and partial ordering is sufficient. Hence, consensus is not required, and the payment system can be realized deterministically in an asynchronous network setting.  \textit{Flash}~\cite{lewis2023flash} is built by encoding payment transactions into the blocks of the blocklace (see Cordial Miners).
The resulting ordering is only partial and hence weaker than the total ordering achieved by Cordial Miners. 

\textit{Mysticeti}~\cite{babel2023mysticeti} improves Cordial Miners for a partially synchronous network. Compared to its predecessor, this consensus protocol allows pipelined leader blocks to be used, improving the finality time for average transactions and faster confirmation in the presence of crashed nodes. It integrates fast payment for owned object transactions using the same underlying DAG and enables checkpoints and epoch-closing mechanisms embedded into the DAG.

\textit{Mahi-Mahi}~\cite{Mahi-Mahi} is a practical improvement of Cordial Miners for an asynchronous network. Like Mysticeti, this protocol allows multiple leaders to be committed per round. In addition, Mahi-Mahi developed the commit rule of Cordial Miners, which resulted in an improved average latency to commit a leader.

\subsection{Consistency-focused Protocols with Certified DAG}\label{consensus_RB}
These consensus protocols are based on classical BFT algorithms. These protocols are round-based, i.e., a node can increase the round number only when the DAG contains a quorum of blocks with the current round number.  These protocols optimize their performance by assigning leaders for certain rounds and using special commit rules for leader blocks. Committed leader blocks form a backbone sequence that allows for the partitioning of the DAG into slices and the deterministic sequencing of blocks in the slices.

\textit{Aleph}~\cite{gkagol2019aleph} protocol improves the Hashgraph complexity by building a round-based structured DAG and employing an efficient binary agreement protocol~\cite{cachin2000random}. Like the Hashgraph protocol, Aleph separates the network layer (communication DAG) from the protocol logic (virtual voting and ordering). Every block created in Aleph is disseminated using a Byzantine Reliable Broadcast (BRB) primitive. Thereby, nodes build a certified DAG. The protocol operates in an asynchronous network and utilizes a trustless ABFT Randomness Beacon to circumvent the FLP-impossibility result, see \cite{FLP}, to reach a total ordering.

 \textit{DAG-Rider}~\cite{keidar2021all} is an asynchronous Byzantine Atomic Broadcast (BAB) protocol. DAG-Rider creates the round-based structured DAG similar to Aleph.
 DAG-Rider is constructed in two layers: communication and ordering layer. In the communication layer, nodes reliably broadcast their proposals (messages) and form a DAG of the messages they deliver. In each round of DAG-Rider, each node broadcasts at most one message, which should contain references (strong edges) to messages (i.e., at least $2f+1$) of previous rounds. The message can also have references (weak edges) to messages of rounds before the previous round. The weak edges ensure the validity property of BAB. Furthermore, in the ordering layer of DAG-Rider, each node observes its local DAG and locally orders all the messages in DAG by employing randomization. This randomization is achieved through a global perfect coin, implemented using threshold signatures that circumvent the FLP-impossibility result.

\textit{Narwhal} \& \textit{Tusk}~\cite{danezis2022narwhal}: Narwhal is a DAG-based, structured mempool incorporating concepts from reliable broadcast~\cite{bracha1985asynchronous} and reliable storage~\cite{abd2005fault}, with the addition of a byzantine fault-tolerant threshold clock~\cite{ford2019threshold} for round advancement. The Tusk protocol is built atop Narwhal to achieve asynchronous consensus with minimal latency. This approach empowers each node to reach a consensus on agreed-upon values by examining its local DAG without additional messages.
      
Tusk refines DAG-Rider, transforming it from theory into an implementable system. This is achieved through three pivotal steps: Firstly, Tusk adopts a Quorum-based reliable broadcast instead of the conventional reliable broadcast utilized in DAG-Rider. Secondly, refining the commit rules enhances the latency in common cases. Lastly, Tusk removes the weak links of DAG-Rider, enabling efficient garbage collection.

\textit{Bullshark}~\cite{spiegelman2022bullshark} is a zero-overhead BFT protocol on top of the Narwhal's DAG optimized for the synchronous case and achieves substantially reduced latency compared to both Tusk and DAG-Rider.  In the partially synchronous version, Bullshark is the most performant and robust compared to existing protocols. Bullshark upholds all the desired properties of DAG-Rider while reducing the common-case latency during the period of synchrony. The main feature of the Bullshark protocol is to exploit the synchronous periods and remove the complex processes of view-change and view-synchronization. Bullshark attains amortized complexity and addresses the imperative facet of fairness. \textit{Shoal}~\cite{spiegelman2023shoal} reduces the latency of non-leader blocks by interleaving two instances of Bullshark.

\textit{BBCA-Chain}~\cite{malkhi2023bbca} introduces a new broadcast primitive, called Byzantine Broadcast with Complete-Adopt (BBCA) to reduce this latency and simplify the consensus logic. BBCA is applied only for leader blocks, while Best-Effort Broadcast (BEB) is applied for all other blocks. The nodes build the DAG as a mixture of certified and optimistic approaches. In addition, BBCA-Chain makes all rounds symmetric by assigning leaders every round. Note that there has been a similar effort; specifically, \textit{Sailfish}~\cite{shrestha2024sailfish} and \textit{Shoal++}~\cite{arun2024shoal++} assign a leader node for every round and allow committing even before BRB instances deliver voting blocks. 

\section{Security and Attack Vectors} \label{security-attack}

Consensus protocols must satisfy two essential properties: \textit{safety} and \textit{liveness}. \textit{Safety} ensures that all honest nodes agree on a consistent state, free of conflicting information, while \textit{liveness} guarantees the continuous addition of new transactions to the ledger.

Security assessments of consensus protocols are conducted under specific theoretical assumptions, primarily defined by the network model (see Section~\ref{sec:networkModel}) and the adversary model, which is typically described by the proportion of faulty or malicious nodes or weights in the system.

In academic analysis, security properties are proven to hold under these assumptions. Often, specific attack vectors reveal protocol vulnerabilities and help establish tight security bounds, highlighting optimal adversarial strategies. Given the varying theoretical approaches to analyzing protocol security, we categorize only the main attack vectors relevant to DAG-based consensus in the following section.

\subsection{Attack Vectors}

We focus here on attack vectors specific to the DAG structure in consensus protocols, excluding general attacks like Sybil, DDoS and Spamming, and double-spending attacks, common across blockchain systems.

DAG-based protocols are susceptible to attacks that leverage the following adversarial actions:
\begin{itemize}
    \item Interfering with network communication (as constrained by the network model) and withholding blocks selectively.
    \item Sending blocks to only a subset of honest nodes.
    \item Producing blocks beyond the protocol’s allowed rate (relevant in PoS, committee-based, and permissioned systems, but generally not in PoW protocols).
\end{itemize}

We categorize attacks by protocol focus—\textit{availability-focused} or \textit{consistency-focused}—and by DAG structure (structured vs. unstructured) and certification type (optimistic vs. certified). Availability-focused protocols are generally more vulnerable to safety attacks, while consistency-focused protocols often encounter liveness issues.

Attacks are valuable in theoretical analysis, as they help determine security bounds, and understanding their impact ensures better security resilience. Below, we summarize several key attack types relevant to DAG-based structures:

\textit{Balance Attack}: A balance attack aims to keep a distributed ledger system in an undecided or "balanced" state, where the network is split between at least two competing subDAGs. This can lead to different outcomes depending on the protocol type: for consistency-focused protocols, the attack causes \textit{liveness} issues by stalling confirmation; for availability-focused protocols, it creates \textit{safety} issues by risking conflicting views.

The balance attack on Conflux \cite{li2020decentralized} demonstrates this vulnerability under high throughput conditions \cite{yu2020ohie}. By maintaining balanced subDAGs, adversaries can manipulate the protocol’s confirmation rules to induce either liveness or safety issues based on the timing and visibility of the attack.

GHOST, although secure at low mining rates \cite{kiayias2019trees}, is susceptible to balance attacks when mining rates increase \cite{Kiffer:18, natoli2016balanceattackproofofworkblockchains}, effectively limiting its throughput in a manner similar to Bitcoin. Protocols that rely on GHOST, such as Inclusive \cite{lewenberg2015inclusive} and Conflux \cite{li2020decentralized}, inherit this limitation. 

Recent advancements address this challenge in different ways. GHOSTDAG \cite{sompolinsky2021phantom} introduces a parameter to handle high throughput, enhancing resilience against balance attacks, while DAG-Knight \cite{sompolinsky2022dag} offers an adaptive solution by dynamically adjusting protocol parameters based on the system’s throughput.

\textit{Parasite-Chain Attack}: In this attack, the adversary constructs a hidden subDAG in parallel to the honest DAG, intending to replace it at some point. This is particularly effective in protocols like IOTA’s Tangle and SPECTRE, which rely on recent tip selection strategies \cite{popov2018tangle, sompolinsky2016spectre}.

\textit{Equivocation Attack}: Equivocation involves presenting different information to different nodes, leading to conflicting states. This attack is challenging for deterministic systems that aim for linearizability, as demonstrated in foundational work on Byzantine fault tolerance \cite{lamport1982byzantine, chun2007attested}. Every PoS or permission system is concerned, and there is a natural connection to the balancing attack.

\textit{Liveness Attack}: Liveness attacks target the confirmation or finalization of transactions by selectively slowing down consensus.  Leader-based DAG protocols face liveness risks from attacks that delay leader block broadcasts, e.g., \cite{Mahi-Mahi}, or introduce equivocation of the leader blocks, both of which disrupt consensus progression.

\textit{Censorship Attack}: Censorship attacks aim to exclude specific blocks or transactions by withholding blocks from particular nodes, hindering consensus progress. In Avalanche, for example, selective censorship can cause liveness issues, see~\cite{whenSpringIsComing}, although recent work has proposed mitigations \cite{rocket2019scalable, buchwald2024frostybringingstrongliveness}.

\textit{Fork-Bomb Attack}: Initially described in \cite{gkagol2019aleph}, this spam attack forces honest nodes to process excessive data, risking system overload. 

\textit{Data-Availability Attack}: In this attack, adversaries withhold blocks from certain parts of the network, risking protocol stalling \cite{gkagol2019aleph}.

Table \ref{tab:attackDAG} provides an overview of these attack vectors across different DAG-based protocols, categorizing protocols analyzed under each type of attack. This categorization does not imply vulnerability but indicates that the protocol has been analyzed for resilience against these attacks. Additionally, the presence of a protocol in a specific category suggests that the respective attack vector may be relevant or of interest, given the protocol’s design.

We also note the absence of attacks for certified DAGs. This stems from the additional use of certificates and reliable broadcast primitives. These external certificates present a possible attack vector, as the block signature flood attack for Tusk studied in \cite{DoSResilience}. 

\begin{table*}[!htbp]
\centering
\caption{Studied attack vectors in DAG-based DLTs}
\label{tab:attackDAG}

\begin{tabular}{|c|c|c|c|c|}
\hline
\multirow{2}{*}{} & \multicolumn{2}{c|}{Availability-Focused} & \multicolumn{2}{c|}{Consistency-Focused} \\
\cline{2-5}
 & Structured DAG & Unstructured DAG & Optimistic DAG & Certified DAG \\
\hline
\multirow{1}{*}{Balance Attack} & \cite{lewenberg2015inclusive, li2020decentralized, yu2020ohie, GHAST, wang2023security} & \cite{popov2018tangle, churyumov2016byteball, popov2020coordicide, kovalchuk2020upper, muller2022tangle} &   &  \\
\hline
\multirow{1}{*}{Parasite-Chain Attack} &  &  \cite{popov2018tangle,sompolinsky2016spectre}  &  &  \\
\hline
\multirow{1}{*}{Equivocation Attack} & \cite{wang2023security} &   & \cite{baird2020hashgraph, xiang2021jointgraph, keidar2022cordial, babel2023mysticeti, chursin2024adeliedetectionpreventionbyzantine}  &  \\
\hline
\multirow{1}{*}{Liveness Attack} &  &  \cite{feedbackcontrol, MitigationLiveness, whenSpringIsComing}  &  \cite{Mahi-Mahi} &  \\
\hline
\multirow{1}{*}{Censorship Attack} &  & \cite{feedbackcontrol, MitigationLiveness, whenSpringIsComing}  &   &  \\
\hline
\multirow{1}{*}{Fork-Bomb Attack} & \cite{gkagol2019aleph} & \cite{gkagol2019aleph}  & \cite{gkagol2019aleph, chursin2024adeliedetectionpreventionbyzantine}   &  \\
\hline
\multirow{1}{*}{Data-Availability Attack} &\cite{gkagol2019aleph}  & \cite{gkagol2019aleph}  & \cite{gkagol2019aleph}  &  \\
\hline
\end{tabular}
\end{table*}

\section{Desirable Properties} \label{desirable-properties}

\subsection{Ordering}
Ordering in distributed ledgers refers to the organization of vertices (blocks/transactions). In blockchain, blocks are typically organized linearly, while in DAG-based systems, ordering depends on factors like graph structure and the underlying consensus protocol. Ordering in DAG-based DLTs can be classified into two types: Partial order and Total order.

\textit{Partial order} arranges DAG blocks through topological sorting, common in protocols like IOTA~\cite{popov2018tangle}, Graphchain~\cite{boyen2018graphchain}, SPECTRE~\cite{sompolinsky2016spectre}, and Nano~\cite{lemahieu2018nano}, where not all blocks are reachable from each other. This approach enables efficient handling of simple payment transactions but limits the use of smart contracts, as shared global states are not inherently supported.

\textit{Total order} organizes blocks in a linear sequence based on parameters such as weight, votes, or timestamp. Some protocols, like GHOST~\cite{sompolinsky2015secure}, PHANTOM~\cite{sompolinsky2021phantom}, and DAG KNIGHT~\cite{sompolinsky2022dag}, determine total order as blocks are added to the DAG by constructing a backbone chain. Others, including the consistency focused protocols, calculate order after rounds of leader selection, which leads to a total order. In Slipstream~\cite{Slipstream}, total ordering is achieved during synchronous network conditions without a dedicated leader.

\textit{Fast Path:} To optimize transaction processing, some DAG-based systems combine partial and total ordering. Simple payment transactions, which do not require global consensus, are confirmed through partial ordering for rapid verification. In contrast, transactions needing a consistent global state, such as those in shared smart contracts, use total ordering.

Sui-Lutris~\cite{blackshear2023sui}, Mysticeti-FPC~\cite{babel2023mysticeti}, and Slipstream~\cite{Slipstream},  implement this approach with variations. Sui-Lutris includes a subprotocol allowing nodes to confirm owned-object transactions \textit{before} they are included in consensus blocks by constructing explicit transaction certificates. In both Mysticeti-FPC and Slipstream, confirmation of owned-objects and UTXO transactions happens \textit{after} including them in consensus blocks by interpreting the local DAGs.

\subsection{Fairness}
Fairness is a fundamental consideration in consensus protocols within DLT systems, addressing technical, economic, and social dimensions. Unfair conditions can foster dissatisfaction among participants, potentially slowing technology adoption. For DAG-based protocols, fairness often involves accepting contributions from slower but honest nodes, defined as those with lower computing power in PoW-based protocols, lower stakes in PoS-based protocols, or substantial communication delays.

In traditional blockchain protocols, centralized, leader-based designs often challenge fairness where a single leader per round has disproportionate influence over block ordering and transaction inclusion. By contrast, DAG-based consensus protocols distribute the role of advancing the ledger across multiple nodes, reducing individual influence over state progression and enhancing fairness.

Raikwar et al.~\cite{raikwar2023fairness} examine fairness in DAG-based DLTs, focusing on participant inclusion, consensus ordering, and component roles within the protocol. They define various fairness criteria and propose methods to support fair participation across different aspects of the system.
We discuss two important topics of fairness in the following two sections. 

\subsection{MEV Protection}
Ordering in a consensus protocol can be exploited through Miner Extractable Value (MEV) attacks, where an adversary attempts to include, exclude, or reorder clients' transactions to maximize profit. MEV, initially introduced as a measure in Flash Boys 2.0~\cite{daian2020flash}, was later renamed Maximal Extractable Value to reflect its broader applicability. These attacks allow validators to extract value at the cost of regular users, impacting the fairness and efficiency of the protocol.

In major DLTs like Ethereum, the urgency for MEV protection is well-known. There have been a few studies to countermeasure these MEV attacks. Yang et al.~\cite{yang2022sok} present a SoK on MEV attacks and its countermeasures. A recent paper~\cite{MEV-machinelearning} presents important insights about MEV extraction on Ethereum through MEV bot bidding strategies where a bid is the highest bribe per computation. 

MEV attacks can be executed on leader-based DAG protocols, yet they can also occur in unstructured DAG protocols where adversaries can act as block proposers. Fino~\cite{malkhi2022maximal} protocol integrates MEV protection into a BFT-based DAG protocol by employing $k$-out-of-$n$ secret-sharing technique. Nasrulin et al.~\cite{nasrulin2023accountable} present an accountable base layer protocol, L{\o}, which detects and mitigates transaction manipulations by creating a secure mempool of verifiable transactions. 

\subsection{Garbage Collection}
In DAG-based consensus protocols, ensuring validity and fairness demands nodes to have unlimited (fast) memory, posing challenges for deploying these protocols. Due to the necessity for boundless memory, nodes can't garbage collect old data, risking the loss of honest but slower nodes' blocks. Garbage collection clashes with fairness, especially within an asynchronous network framework where blocks may experience indefinite delays. This creates a crucial trade-off between ensuring fairness and optimizing performance.

Garbage collection methods vary across DAG-based consensus protocols, aiming to balance the need for memory optimization. The blocks of the DAG can be garbage collected based on their depth (or round), timestamp (or age), or finality. It can also be performed in DAG-based consensus protocols Vite~\cite{liu2018vite} and Jointgraph~\cite{xiang2021jointgraph}, which maintain snapshot blocks (or chain). Consequently, older blocks can be garbage collected since their history is stored within the snapshot blocks. Jointgraph uses its snapshot and storage events to release the previous memory.

Narwhal~\cite{danezis2022narwhal} cleans up its DAG by discarding information up to a specific round from the genesis. However, this compromises fairness at the block level, risking disposal of slower nodes' data before proper ordering. In contrast, Bullshark~\cite{spiegelman2022bullshark} employs a garbage collection mechanism while mitigating fairness issues. Bullshark and  IOTA 2.0~\cite{iota} garbage collects blocks that are not committed for a long period of time, ensuring fairness post-GST during synchronous periods.

\section{Discussion} \label{discussion}

\subsection{Broadcast Techniques}
Broadcast techniques in DAG-based consensus protocols ensure the propagation of blocks. The protocols using a certified DAG, see Table \ref{tab:consensusDAG},  rely on underlying reliable and consistent broadcast primitives 
to disseminate blocks, whereas the protocols using an optimistic DAG (this includes, in addition, most availability-focused protocols, see Table \ref{tab:consensusDAG}) use best-effort broadcast. 
\begin{itemize}[leftmargin=*]
\item \textit{Byzantine Consistent Broadcast (BCB)} guarantees that a sender cannot send conflicting blocks ( equivocate). It employs a multi-step communication pattern between the sender and the receiver. 
In DAG-based protocols, e.g. Tusk~\cite{danezis2022narwhal}, BCB is treated as an abstract entity.  For each block, the block creator calls its own BCB broadcast instance. In \cite{stathakopoulou2023bbca,malkhi2023bbca}, an abortable variant of BCB, called BBCA, is introduced that is then used to broadcast only the leader blocks. 
    
\item \textit{Byzantine Reliable Broadcast (BRB)} ensures the reliable dissemination of messages, even when a certain number of nodes in the system may be malicious or faulty. BRB requires at least $f+1$ honest nodes to unanimously agree on a value $v$ before committing across all nodes. While BRB adds an extra step compared to BCB, DAG-based protocols, e.g., Bullshark~\cite{spiegelman2022bullshark}, use BRB to ensure the liveness of the protocol.

\item \textit{Best-Effort Broadcast (BEB)} facilitates block dissemination optimistically. In this approach, when a node intends to share a block, it simply transmits it with some (potentially unknown) part of its causal history to all other nodes, requiring only a single round of communication instead of at least two rounds when utilizing CB. This approach is demonstrated in Hashgraph~\cite{baird2020hashgraph} and Cordial Miners~\cite{keidar2022cordial}. To minimize the communication overhead, many actual implementations of DAG-based protocols do not send the causal history of the block; instead, they use a pull strategy to minimize the communication overhead in the common case.
\end{itemize}

\subsection{Latency and Throughput}
Network latency and round trip time (RTT) are essential for consensus. Latency is the time for a packet to travel from point A to point B, while RTT encompasses the time for a packet to go from A to B and back, including encoding, queuing, processing, decoding, and propagation delays. These factors, typically consistent for a given pair of endpoints, can be affected by network congestion, adding variability to RTT. It's important to note that latency is not necessarily half of RTT due to potential differences in delays between endpoints.

Consistency-focused consensus protocols based on BFT incur high latency in asynchronous networks. Recent works aim to minimize the latency of these protocols. Spiegelman et al.~\cite{spiegelman2023shoal} introduced Shoal, a protocol-agnostic framework that improves the latency of the BFT-based consensus protocols. Shoal~\cite{spiegelman2023shoal} enhances the latency of Narwhal-based consensus protocols by employing a leader reputation mechanism to prevent failures and introducing pipelining to ensure a well-ordered DAG construction without timeout requirements. 

Liu et al.~\cite{liu2023reducing} present a novel DAG-based commit rule that effectively reduces transaction latency in the UTXO model. Their novel commit rule accelerates transaction confirmation and reduces the confirmation latency. Another recent work~\cite{liu2023flexible} presents a flexible advancement in asynchronous BFT consensus protocols, bridging the ordering and agreement components and reduces the latency of the consensus.

The primary advantage of DAG-based protocols over blockchains is their enhanced throughput. Various DAG-based consensus protocols demonstrate high throughput.
Amores-Sesar and Cachin~\cite{amores2023we} present a construction that takes a DAG-based consensus protocol $\Pi$ as input and outputs a new DAG-based protocol ${\Pi}'$ which has superior throughput and latency. 

\subsection{Execution Bottlenecks in High-Throughput DAG-Based DLTs}
In DAG-based DLTs, the increased throughput shifts the primary bottleneck from data writing and consensus to transaction execution, necessitating an efficient, parallelized execution layer. Prism addresses this by decoupling consensus from transaction validation, achieving high throughput independently of consensus finality \cite{prismexecutionbottleneck}. This design emphasizes the need for scalable execution frameworks. Block-STM \cite{gelashvili2022blockstmscalingblockchainexecution} scales smart contract processing through speculative execution, dynamically resolving conflicts to maintain high throughput. Pilotfish \cite{kniep2024pilotfishdistributedtransactionexecution}, a distributed execution engine, leverages a novel crash-recovery protocol and versioned-queues scheduling to ensure state consistency while maximizing parallelism, even with complex read-write dependencies.

\subsection{Mathematical Models and Simulations}
A mathematical model can analyze DAG-based consensus protocol metrics, but constructing a generic model to simulate and evaluate these protocols is a challenging task. Some attempts have been made to build such models and simulations.

S. Popov's initial mathematical model for a DAG-based protocol~\cite{popov2018tangle} assumed new message arrivals following a unique Poisson process, leading to a conjecture about the stationary rate of unapproved transactions. Simulations supporting this conjecture with homogeneous delays spurred interest in developing new mathematical models for DAG-based protocols, as seen in studies such as~\cite{kusmierz2019properties, park2019performance}. Works like~\cite{li2020direct} explored non-Poisson message arrival scenarios, often assuming a central node managing ledger records with other nodes accessing the ledger state through it. Penzkofer et al.~\cite{penzkofer2021impact} extended Popov's model, introducing heterogeneous delays instead of homogeneous ones. Zander et al. introduced DAGsim~\cite{zander2019dagsim}, a multi-agent simulator based on a heterogeneous delay model for DAG-based protocols. Recent research~\cite{kumar2022effect, mueller2023stabilitylocaltippool} employed a discrete-time Markov chain to model the evolution of unapproved transactions under heterogeneous delay. Lin et al. developed TangleSim~\cite{lin2023tanglesim} to implement leaderless Tangle 2.0 in various network scenarios and byzantine environments.
Bullshark~\cite{spiegelman2022bullshark}.

For the simulation of deterministic protocols, Schett and Danezis~\cite{schett2021embedding} formalized a blockDAG protocol implementing a reliable point-to-point channel, embedding deterministic BFT protocols while maintaining safety and liveness properties. They utilized deterministic state machines for node communication, affirming properties claimed by Hashgraph~\cite{baird2020hashgraph}, Blockmania~\cite{danezis2018blockmania}, and Flare~\cite{rowan2019flare} within their framework. Attiya et al.\cite{attiya2023faithful} extended this work, faithfully simulating non-deterministic (Randomized) BFT protocols on blockDAG, capturing probabilistic guarantees and enabling analysis of protocols like Aleph\cite{gkagol2019aleph}, DAG-Rider~\cite{keidar2021all}, and Bullshark~\cite{spiegelman2022bullshark}.

We want to mention that a series of recent performance evaluations are based on similar testbeds. For instance, the performance of optimistic DAGs is evaluated in \cite{babel2023mysticeti} (Mysticeti), \cite{Mahi-Mahi} (Mahi-Mahi),  and in~\cite{obelia} (Obelia). The comparison of certified DAG-based consensus protocols such as Bullshark, Shoal, and Sailfish is provided in~\cite{shrestha2024sailfish}.

\subsection{Lightweight DAG-based Protocols}
DAG-based protocols provide advantages in performance and scalability over traditional blockchains but encounter challenges related to storage scalability due to data redundancy. This issue increases costs for developers of decentralized applications (dApps), who must efficiently track and verify state changes. This is a general challenge for high-throughput DLTs, and here we provide a brief overview of solutions specifically proposed for DAG-based protocols.

To support diverse applications like Vehicular Social Networks (VSN) and the Internet-of-Things (IoT), lightweight protocols have been developed to address resource constraints. Yang et al.~\cite{yang2020ldv} proposed LDV, a DAG-based protocol optimized for VSNs, which selectively retains necessary information and prunes historical data, making it suitable for resource-limited vehicles. Similarly, Cherupally et al.~\cite{cherupally2021lightweight} introduced LSDI, a scalable DAG-based ledger designed for verifying IoT data integrity within cloud-based architectures. LSDI enhances lightweight functionality by pruning outdated segments.

Recently, Dai et al.~\cite{dai2023geckodag} proposed GeckoDAG, a lightweight DAG protocol that addresses data redundancy issues. GeckoDAG reduces account and reference redundancy by consolidating transactions and introducing \textit{reference override} for efficient reference management. These optimizations reduce storage requirements without compromising security.


\section{Future Research} \label{futureResearch}

The following presents an overview of emerging discussion points and future research questions. 

\textit{Performance Evaluation:} No standardized benchmarks exist for evaluating DAG-based consensus protocol performance. Research papers often focus on specific metrics, emphasizing protocol benefits. Developing standardized performance metrics is crucial for objectively evaluating whether DAG-based approaches mitigate performance bottlenecks in chain-based protocols. Actual protocol implementations, when available, frequently diverge from their theoretical designs due to practical considerations. This discrepancy underscores the importance of real-world testing in validating the effectiveness of DAG-based systems.

\textit{Security Analysis:} Formal security proofs for DAG-based protocols appear sporadically and vary in scope and depth. Developing a comprehensive framework is necessary to understand and compare these protocols' security features. Furthermore, the confluence of ideas and concepts within these protocols highlights the need for a generalized abstraction. This abstraction would facilitate a universal treatment of security measures and attack vectors, allowing for a more structured analysis of protocol security.

\textit{Incentive Attacks and Game-Theoretic Challenges:}  In DAG-based distributed ledgers, traditional blockchain incentive concerns—such as protocol deviations for selfish gain—remain underexplored.  Analyzing these challenges in the context of availability- and consistency-focused protocols could reveal generalized principles for ensuring fair participation. For instance, availability-focused systems might require incentive mechanisms that prevent balancing attacks, while consistency-focused systems may benefit from mechanisms that discourage strategic manipulations of block ordering and equivocations.

\textit{Transaction duplicates:}
Transaction duplicates pose a challenge for DAG-based protocols to enhance system efficiency. The issue arises when multiple validators pull transactions from a shared mempool for parallel block proposals, risking transaction duplication and negating parallel execution benefits. A solution like assigning transactions to specific validators could mitigate repetition but potentially enable censorship and centralized control, undermining decentralization principles.
Some DAG-based projects address it through ad hoc methods as the random association between transactions and nodes. However, there is a need for research focused on strategies to prevent transaction duplication without sacrificing system fairness or decentralization.

\textit{Fairness:} 
Research on fairness within DAG-based protocols should assess the implications of more open writing access on transaction equity, particularly in the Maximal Extractable Value (MEV) context. It is essential to consider how the total ordering of transactions, despite more open writing access, can influence MEV opportunities. This inquiry should extend to the role of entry points, the underlying peer-to-peer (P2P) and Remote Procedure Call (RPC) nodes, which are critical in MEV dynamics. Investigating these aspects will provide insight into whether decentralization in writing access successfully mitigates fairness issues. Moreover, it is essential to understand the relationship between access control, tokenomics, and consensus mechanisms to understand fairness and evaluate the level of decentralization.

\textit{Privacy:} DAG-based consensus protocols should prioritize constructing privacy-preserving mechanisms that leverage higher bandwidth in DAG structures. Implementing privacy in DAG-based protocols introduces challenges. The tip selection process may become costlier as validating transactions in the tips requires computational effort due to cryptographic privacy measures on transaction data encoded within the tips. This, alongside concerns about transaction ordering and transaction auditability in the DAG, underscores the significance of selecting the appropriate privacy technique.
Therefore, efforts should be directed at developing protocols that protect identities and transaction data without sacrificing performance, exploring ways to balance confidentiality with the enhanced capacity of the DAG network.

\textit{Mempool Abstraction:} Mempool abstraction allows decoupling transaction dissemination from the consensus protocol, streamlining the ordering process for better efficiency. The selection of transactions can be optimized to prevent redundancy upon their block inclusion. Additionally, leveraging load-balancing protocols for distribution across nodes can contribute to equitable transaction handling. Consequently, further investigation is required to explore how mempool abstraction affects fairness within DAG-based consensus protocols and its implications for the cost of transaction ordering.

\section{Conclusion} \label{conclusion}
In this SoK, we analyzed DAG-based consensus protocols, classifying them by availability and consistency requirements. We outlined key models, components, security aspects, potential attack vectors, and relevant countermeasures. The study also highlighted recent developments in execution scalability and fairness, reflecting the unique demands of high-throughput DLTs.

\section{Call for Contributions and Future Updates}
While we have aimed to provide a comprehensive overview of relevant works in DAG-based consensus protocols, some research may have been unintentionally omitted. We encourage researchers to contact us if any important work has been missed. We also intend to keep this study up-to-date as long as there is interest. We invite you to share new developments with us or cite this work to facilitate discovery.

\section{Acknowledgement}
Mayank Raikwar has been supported by IOTA Ecosystem Development grant.

\bibliographystyle{IEEEtran} 
\bibliography{reference}

\end{document}